\documentclass[twocolumn,showpacs,preprintnumbers,amsmath,amssymb]{revtex4}
\usepackage{graphicx}% Include figure files
\usepackage{dcolumn}% Align table columns on decimal point
\usepackage{bm}% bold math

\begin{document}

%\preprint{APS/123-QED}

\title{Extraordinary transmission caused by symmetry breaking}% Force line breaks with \\
\author{Dan Hu$^{1, 2}$,  Chang-Qing Xie$^{3}$, Ming Liu$^{3}$, and Yan Zhang$^{2}$}
\email{*yzhang@mail.cnu.edu.cn} %% email address is required
\affiliation{$^{1}$Department of Physics, Harbin Institute of Technology, Harbin, 150001 China \\
$^{2}$Beijing Key Lab for Terahertz Spectroscopy and Imaging, Key
Laboratory of Terahertz Optoelectronics, Ministry of Education,
and Department of Physics, Capital Normal University, Beijing, 100048 China\\
$^{3}$Laboratory of Micro-processing and Nanotechnology, Institute
of Microelectronics, Chinese Academy of Sciences, Beijing, 100029
China}

\date{\today} % It is always \today, today,
             %  but any date may be explicitly specified

\begin{abstract}
The terahertz transmission properties of symmetrical and asymmetrical annular apertures arrays (AAAs) are investigated both experimentally and numerically. It is found that only odd order resonant modes are observed for the symmetrical structures but both odd and even order resonances can be shown for the asymmetrical structures. Breaking the symmetry of  AAAs  by gradually displacing the H-shaped AAAs to U-shaped AAAs
allows an intensity modulation depth of $99 \%$ of the second order
resonance. Simulation results verify the experimental conclusions
well. This result provides a tremendous opportunity for terahertz
wavelength tunable filtering, sensing, and near-field imaging.
\end{abstract}

% PACS, the Physics and Astronomy
                             % Classification Scheme.
%\keywords{Suggested keywords}%Use showkeys class option if keyword
                              %display desired
\maketitle

Since the extraordinary transmission (EOT) phenomenon of
sub-wavelength metallic apertures arrays  has been reported by
Ebbesen and his co-workers \cite{Ebbesen98,Barnes99}, it has
been paid extensive attentions to understand the underlying physical
mechanisms \cite{Ghaemi98,Maier07}. Although it is still debated on
the real mechanism of the EOT, surface plasmon polaritons (SPPs)
originating from the coupling of light with the surface charges
oscillation on the metal-dielectric interface \cite{Ghaemi98} and the
localized surface plasmon resonance (LSPR) excited on the hole ridge
are recognized as the main possible contributions to the EOT
\cite{Maier07}. Moreover, it has been shown that the transmission
spectral response of sub-wavelength metallic apertures arrays are
affected by the cell size \cite{Molen04}, cell shape
\cite{Lee06,Lee09}, lattice periodicity \cite{Lee10}, metal film
thickness \cite{Wang09}, the refractive index of the adjacent
medium \cite{Azad05}, as well as the polarization of incident light
\cite{Masson06}. Annular aperture arrays (AAAs) have also been demonstrated to
have EOT \cite{Poujet07,Lu10}. The influence of geometric and
structure parameters of the AAAs on the EOT properties has been
extensively investigated. However, the influences of the
symmetry of the AAAs on the EOT have not been reported. In this
Letter, the transmission properties of the symmetric and asymmetric
 AAAs are investigated both experimentally and theoretically.
The interactions of the terahertz (THz) electromagnetic radiation
with the AAAs have been measured with the traditional THz
time domain spectroscopy (TDS). The field distributions of the
resonant wavelengths are simulated with the finite difference time
domain (FDTD) method. The physical original of the resonant
enhanced transmission has been analyzed. It is found that the
symmetry breaking can enhance the even order mode EOT.

\begin{figure}
\includegraphics[width=8.0cm]{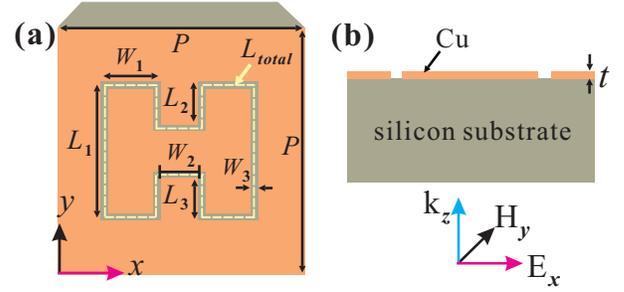}
\caption{\label{fig:sample} Schematic view of the H-shaped AAAs. (a) Top view and (b)
side view of the sample. The x-polarized THz illumination is also
indicated in the figure. }
\end{figure}

The proposed AAAs structure is schematically shown in Fig. \ref{fig:sample}. It is a H-shaped
annular slit. The outer arm length of the H-shaped slit is $L_1$, the upper and lower
inner arm lengths are  $L_2$ and $L_3$, the outer and
inter arm widthes are $W_1$ and $W_2$, respectively.  The slit width is
$W_3=2.5 \mu m$ and the period ($\Lambda $) of the sample is $200 \mu m$. This sample is symmetrical along both $x$ and $y$ directions. The total average circumference of the AAAs, which is the average length of out and inner edge of the slit, is denoted as $L_{total}$. The samples
are fabricated onto a thin copper film ($t=200 nm$) with the
conventional photolithography and metallization processes. The
substrate is a $500 \mu m$ thick doubly polished p-type silicon
wafer.

The transmission spectra of the samples are measured by using a
typical THz TDS. The THz wave pulses are
generated from a dipole-type photoconductive antenna illuminated by
a $100fs$ laser with $800nm$ center wavelength. The valid range of
the system is $0.2-2.6THz$. The detection is achieved by a ZnTe
crystal. For the reference, a bare silicon wafer identical to the
one on which the AAAs are fabricated is used. The
transmitted THz pulses are measured at normal incidence such that
the electric and magnetic field of the incident radiation are in
AAAs' plane. All measurements are done at room temperature and in a
dry atmosphere to mitigate water absorption. The normalized
transmittance is obtained by
$T(\nu)=|E_{sample}(\nu)/E_{ref}(\nu)|$, where $E_{sample}(\nu)$ and
$E_{ref}(\nu)$ are Fourier transformed transmitted electric field of
the sample and reference pulses, respectively.

\begin{figure}
\includegraphics[width=8.0cm]{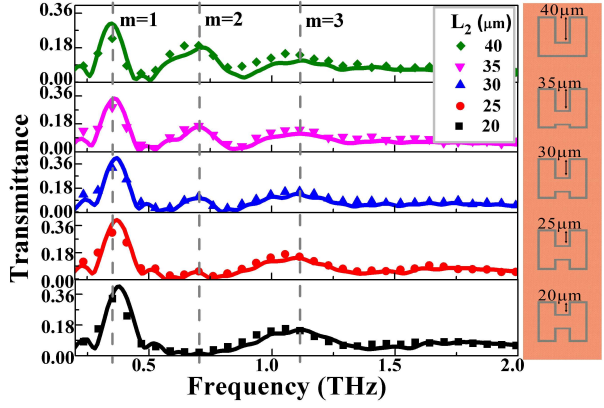}

\caption{\label{fig:experiment} Measured (symbol) and simulated (solid line) transmission spectra for AAAs with same average
circumference $L_{total}$ but  different $L_2$. }
\end{figure}

The transmission spectra of five samples with different $L_2$ are
shown in Fig. \ref{fig:experiment} with symbols. The parameters of the structures are selected as
follows: The widthes $ W_1$ and $W_2$ are $30 \mu m$ and $20
\mu m$, respectively. The length $L_1$ is $80 \mu
m$. The length $L_2$ increases from $20 \mu m$ to $40 \mu m$ with
step of $5 \mu m$ and thus $L_3$ reduces from $20 \mu m$ to $0\mu
m$. The average circumference of the slit is kept as a constant as $L_{total}=390 \mu
m$. For $L_2=L_3=20 \mu m$, the sample is H-shaped AAAs and symmetrical along both $x$ and
$y$ directions. For $L_2=40 \mu m$ and $L_3=0 \mu m$, the sample becomes to U-shaped AAAs and it is only symmetry along the $x$ direction. It can be seen that two resonance peaks at $0.35 THz$ and $1.11 THz$ appear in all transmission spectra.
With $L_2$ increasing, the symmetry of the structure along the $y$
direction has been broken, the locations of these two peaks nearly
unchanged but a new resonance peak appears at $0.70 THz$. The
transmission for $0.70 THz$ is enhanced from $1.97\%$ for symmetrical
structure to $17.5\%$ for asymmetrical one, the intensity modulation
depth is $99\%$.

For the AAAs structure, the localized surface plasmonics will be excited under
the light illumination. The surface charges are assembled at the
edge of the metal. Based on the basic idea of Ref. \cite{Neubrech06}, the
resonance appears when the following condition is satisfied:

\begin{equation}
f_m=\frac{m c} { n_{eff} L_{total}}, m=1, 2, 3,.....
\end{equation}
where $f_m$ is the resonant frequency, $m$ is an integer which is
the order of the resonant mode, $c$ is the speed of light in vacuum,
$L_{total}$ is the average circumference of the AAAs, and $n_{eff}$ is
the effective refraction index for the localized surface plasmonic wave. In our case, $n_{eff}$ is approximatively equal to $(1.0+n_{sub})/2$, $n_{sub}$ is the refraction index of the substrate. Under this condition,
the standing wave of the localized surface plasmon can be generated along the wedges of the
metallic film. Furthermore, the distribution of localized surface plasmon should also satisfy the symmetry requirement of the samples. All samples considered here are symmetrical along the $x$
direction and the illumination light is polarized along the $x$
direction, thus the distribution of the assembled surface charges should
be antisymmetrical along the $x$ direction. If the sample is also symmetrical along the $y$ direction, the distribution of the assembled surface charges should also be even symmetrical along the $y$ direction. When the distribution of the assembled surface charges satisfy both the standing wave condition and symmetry condition,  the corresponding transmission can be greatly enhanced. For all samples considered here, the odd mode (for example, $m=1$ or $m=3$) can easily satisfies these requirements no matter the sample is symmetrical along the $y$ direction or not. However, if the resonant mode is even, the symmetry of the sample along the $y$ direction will affect the distribution of the surface charges and hence the transmission amplitude. It requires the distribution of the surface charges
is also symmetrical along the $y$ direction. Thus, the
standing wave condition for second mode can not be stratified any more, therefore, the transmission for the second resonant mode will be greatly reduced, as shown in Fig. 2 with $L_2=20 \mu m$. But when the
symmetry along the $y$ direction is broken, the requirement of the
symmetry of surface charge distribution along this direction does
not exist, the standing wave of the localized surface plasmon can be
generated and the resonant peak appears, as shown in Fig. \ref{fig:experiment}
with $L_2=40 \mu m$.

In order to get an insight into the nature of the resonant
transmission for the proposed structures, the numerical simulations
are carried out by using the FDTD approach \cite{Simulations} . The
metal copper is described by the perfect electric conductor to fit
its realistic characteristic in the terahertz range. The
refraction indices of the air and substrate silicon are set as 1.0
and 3.31 in the simulations, respectively. The calculated
transmission spectra are also shown in Fig. \ref{fig:experiment} with solid lines. It
can be found that the simulation results correspond to the
experiment measurements well. From the simulation results, it can also be
found that the sample with $L_2=20 \mu m$ has two resonant
peaks at $0.35 THz$ and $1.11 THz$, which correspond to $m=1$ and $m=3$, respectively. But the sample with $L_2=40 \mu m$ has three resonant peaks. A new peak corresponding to $m=2$ appears at $0.70 THz$.
With $L_2$ increasing, the transmission for the second resonant mode enhanced.

\begin{figure}
\includegraphics[width=8.0cm]{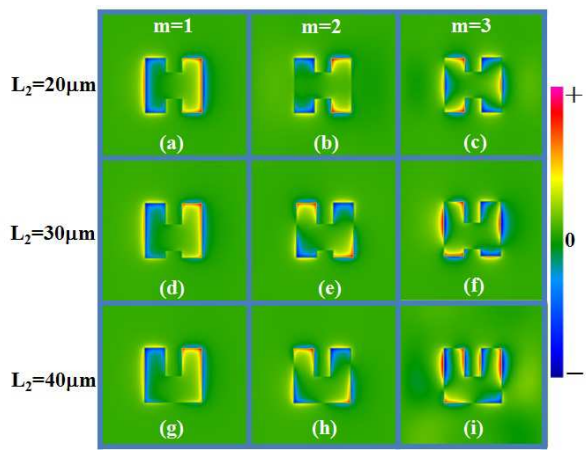}
\caption{\label{fig:simulation} Simulated $E$ distributions on the interface between metal
film and dielectric substrate silicon for three samples at three
resonant wavelengths. The snapshot moment is when the field
amplitude reaches its maximum. Only a unit cell on the x-y plane is
shown. The field quantity is normalized with respect to the incident
field $E_{0}$.}
\end{figure}

Figure \ref{fig:simulation} presents the extremum electric field distributions at the
aforementioned three resonant frequencies for samples with $L_2$
equals $20\mu m$, $30\mu m$, and $40\mu m$, respectively, which
exhibits the distributions of the surface charges. When the structure is symmetrical along both $x$ and $y$ directions, it can be found from Fig.
\ref{fig:simulation}(a)-(c) that the extremum electric field
distributions are antisymmetrical along the $x$ direction but symmetrical
along the $y$ direction. These are natural results for satisfying the symmetrical requirement of the samples. For the odd modes $m=1$ and $m=3$ (correspond to Fig.
\ref{fig:simulation}(a) and (c), respectively), the standing waves are generated along
the edges of the metallic film. One wave crest along the out side of slit in Fig. \ref{fig:simulation}(a) and three crests in Fig. \ref{fig:simulation}(c) can be found.
However, in order to satisfy the symmetry condition along the $y$ direction,  the surface charge wave can not generate a standing wave with mode $m=2$, as shown in Fig. 3(b). Only a temporarily electric field distribution is shown here and no standing wave distribution can be found. For the structure with $L_2=40 \mu m$, the electric field distributions are also antisymmetrical along the $x$ direction. Since the structure is not symmetrical along the $y$ direction any more, the surface charges distributions are not required to be symmetrical along the $y$ direction. There are 1, 2, and 3 standing waves generated for mode $m=1$ (Fig. 3(g)), $m=2$ (Fig. 3(h)) and
$m=3$ (Fig. 3(i)), respectively. The even mode standing wave is
formed and thus the corresponding transmission has been enhanced. It can be
concluded that the resonant peak $m=2$ for the structure with
$L_2=40 \mu m$ is caused by the symmetry breaking along the $y$
direction.

\begin{figure}
\includegraphics[width=8.0cm]{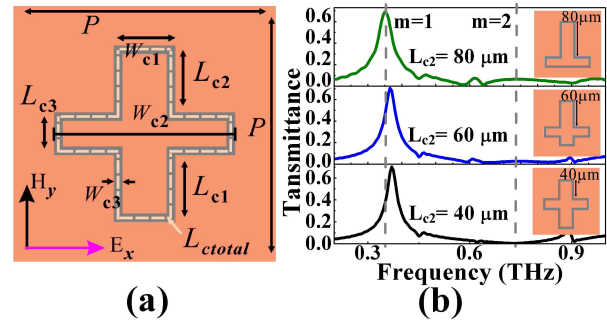}
\caption{\label{fig:cross}(a) Schematic of cross shaped AAAs with fixed parameters: $W_{c1}=30\mu m, W_{c1}=100\mu m, W_{c3}=2.5\mu m, L_{c1}+L_{c2}=80\mu m, L_{c3}=20\mu m, L_{ctotal}=390\mu m,$ and the period $\Lambda=200\mu m$. (b) Simulated spectra for cross shaped AAAs with with same average circumference $L_{ctotal}$ but different $L_{c2}$. }
\end{figure}

For further validation, the transmission spectra for another series of structures are calculated and shown in Fig. \ref{fig:cross}. The AAAs is cross shaped and the average circumference $L_{ctotal}$ is still $390 \mu m$. The girder is shifted from the middle to the bottom of the cross to break the symmetry along the $y$ direction. It can be seen from the transmission spectra that the odd mode $m=1$ appears around $0.35 THz$ for all structures. However, the transmission intensity for second mode $m=2$ increases from $0.39\%$ for symmetrical structure to $6.5 \%$ for the asymmetrical structure. These results also demonstrate that the symmetry breaking can enhance the odd mode resonance.

In summary, the transmission characteristics of the symmetrical and asymmetrical AAAs
with narrow gap are investigated experimentally and numerically. The
measured and simulated transmission spectra reveal that the resonant
transmission are affected by the geometric structure. Both even and
odd order modes can emerge for the asymmetrical AAAs but only odd
order modes appears for the symmetrical AAAs. The appearance of the even
mode is caused by the symmetry breaking. This research indicates a
new freedom in tuning the electromagnetic response, which offers a
path to design more robust plasmonic devices.

%\section*{Acknowledgements}

This work was supported by the 973 Program of China (No.
2011CB301801) and National Natural Science Foundation of China (No.
10904099 and 11174211).

%\newpage
%\centerline{\bf Figure captions}
%
%\begin{description}
%
%\item {Fig. 1}
%Schematic plan of the setup and sample geometry.
%
%\item{Fig. 2}
%Measured transmission spectra for the single metallic slit with
%different width.
%
%\item {Fig. 3}
%Measured phase distribution of the transmission spectra for the
%single metallic slit with different width.
%
%\item{Fig. 4}
%Calculated transmission spectra for the single metallic slit with
%different width.
%
%\item{Fig. 5}
%Calculated phase distributions of the transmission spectra for the
%single metallic slit with different width.

%\end{description}
\end{document}